%Paper: hep-ph/9501243
%From: ramond@phys.ufl.edu
%Date: Mon, 09 Jan 95 11:46:47 -0500

%      Page layout, margins (feel free to change)
\hsize=6.5truein
\hoffset=.3truein
\vsize=8.9truein
\voffset=.1truein
%%%%%%%%%%%%%%%%%%%%%%%%%%%%%%%%%%%%%%%%%%%%%%%%%%%%%%%%%%%%%%%%%%%%%%%%%%%%%
% This is sjnl.tex,  from Doug Eardley's jnl, eqnorder and reforder, %
% but combined, modified and reduced to form     %
%%%%%%%%%%%%%%%%%%%%%%%%%%%%%%%%%%%%%%%%%%%%%%%%%%%%%%%%%%%%%%%%%%%%%%%%%%%%%
%  Define pseudo-12pt fonts
\font\twelverm=cmr10 scaled 1200    \font\twelvei=cmmi10 scaled 1200
\font\twelvesy=cmsy10 scaled 1200   \font\twelveex=cmex10 scaled 1200
\font\twelvebf=cmbx10 scaled 1200   \font\twelvesl=cmsl10 scaled 1200
\font\twelvett=cmtt10 scaled 1200   \font\twelveit=cmti10 scaled 1200
\skewchar\twelvei='177   \skewchar\twelvesy='60
%  Define \...point macros to change fonts and spacings consistently
%\def\twelvepoint{\normalbaselineskip=12.4pt
\def\twelvepoint{\normalbaselineskip=14pt
  \abovedisplayskip 12.4pt plus 3pt minus 9pt
  \belowdisplayskip 12.4pt plus 3pt minus 9pt
  \abovedisplayshortskip 0pt plus 3pt
  \belowdisplayshortskip 7.2pt plus 3pt minus 4pt
  \smallskipamount=3.6pt plus1.2pt minus1.2pt
  \medskipamount=7.2pt plus2.4pt minus2.4pt
  \bigskipamount=14.4pt plus4.8pt minus4.8pt
  \def\rm{\fam0\twelverm}          \def\it{\fam\itfam\twelveit}%
  \def\sl{\fam\slfam\twelvesl}     \def\bf{\fam\bffam\twelvebf}%
  \def\mit{\fam 1}                 \def\cal{\fam 2}%
  \def\tt{\twelvett}
  \textfont0=\twelverm   \scriptfont0=\tenrm   \scriptscriptfont0=\sevenrm
  \textfont1=\twelvei    \scriptfont1=\teni    \scriptscriptfont1=\seveni
  \textfont2=\twelvesy   \scriptfont2=\tensy   \scriptscriptfont2=\sevensy
  \textfont3=\twelveex   \scriptfont3=\twelveex  \scriptscriptfont3=\twelveex
  \textfont\itfam=\twelveit
  \textfont\slfam=\twelvesl
  \textfont\bffam=\twelvebf \scriptfont\bffam=\tenbf
  \scriptscriptfont\bffam=\sevenbf
  \normalbaselines\rm}
%       tenpoint

%      Various internal macros
\def\beginlinemode{\endmode
  \begingroup\parskip=0pt \obeylines\def\\{\par}\def\endmode{\par\endgroup}}
\def\beginparmode{\endmode
  \begingroup \def\endmode{\par\endgroup}}
\let\endmode=\par
{\obeylines\gdef\
{}}
\def\singlespace{\baselineskip=\normalbaselineskip}
\def\oneandahalfspace{\baselineskip=\normalbaselineskip
  \multiply\baselineskip by 3 \divide\baselineskip by 2}
\def\doublespace{\baselineskip=\normalbaselineskip \multiply\baselineskip by 2}
\newcount\firstpageno
\firstpageno=2
%% FOLLOWING LINE CANNOT BE BROKEN BEFORE 80 CHAR
\footline={\ifnum\pageno<\firstpageno{\hfil}\else{\hfil\twelverm\folio\hfil}\fi}
\let\rawfootnote=\footnote              % We must set the footnote style
\def\footnote#1#2{{\rm\singlespace\parindent=0pt\rawfootnote{#1}{#2}}}
\def\raggedcenter{\leftskip=2em plus 12em \rightskip=\leftskip
  \parindent=0pt \parfillskip=0pt \spaceskip=.3333em \xspaceskip=.5em
  \pretolerance=9999 \tolerance=9999
  \hyphenpenalty=9999 \exhyphenpenalty=9999 }
\parskip=\medskipamount
\twelvepoint            % selects twelvepoint fonts (cf. \tenpoint)
\overfullrule=0pt       % delete the nasty little black boxes for overfull box
    % Preprint number at upper right of title page
\def\author                     %  Author(s) name(s)  on title page
  {\vskip 3pt plus 0.2fill \beginlinemode
   \singlespace \raggedcenter \twelvesc}
\def\affil                      % Affiliations (can intermix with \author)
  {\vskip 3pt plus 0.1fill \beginlinemode
   \oneandahalfspace \raggedcenter \sl}
\def\abstract                   % Begin abstract
  {\vskip 3pt plus 0.3fill \beginparmode
   \doublespace \narrower \noindent ABSTRACT: }
\def\endtitlepage               % End title page, begin body of paper
  {\endpage                     %       This subsumes \body
   \body}
\def\body                       % Begin text body;  can be used to end
  {\beginparmode}               % \title, \author, \affil, \abstract,
                                % \reference, or \figurecaption modes

%%\def\subhead#1{                 % Subhead;  NOTE enclose the text in {}
%%  \vskip 0.25truein             % e.g., \subhead{A. History of the Problem}
%%  {\raggedcenter #1 \par}
%%   \nobreak\vskip 0.25truein\nobreak}
\def\subhead#1{                 % Subhead;  NOTE enclose the text in {}
  \vskip 0.1truein             % e.g., \subhead{A. History of the Problem}
  {\raggedcenter #1 \par}
   \nobreak\vskip 0.1truein\nobreak}
\def\refto#1{$|{#1}$}           % For references in text as superscript
\def\references                 % Begin references -- basic format is Phys Rev
  {\subhead{References}         % I.e., volume, page, year (space after
%%commas).
   \beginparmode
   \frenchspacing \parindent=0pt \leftskip=1truecm
   \parskip=8pt plus 3pt \everypar{\hangindent=\parindent}}
\gdef\refis#1{\indent\hbox to 0pt{\hss#1.~}}    % Ref list numbers.
\gdef\journal#1, #2, #3, 1#4#5#6{               % Journal reference.  Comma
%%sets
    {\sl #1~}{\bf #2}, #3, (1#4#5#6)}           % off: name, vol, page, year
\def\refstylenp{                % Nucl Phys(or Phys Lett) ref style: V, Y, P
  \gdef\refto##1{$^{(##1)}$}                                % Reference in text
[]
  \gdef\refis##1{\indent\hbox to 0pt{\hss##1)~}}        % Ref list numbers)
  \gdef\journal##1, ##2, ##3, ##4 {                     % Journal reference
     {\sl ##1~}{\bf ##2~}(##3) ##4 }}
\def\refstyleprnp{              % Input like pr, output like np!!
  \gdef\refto##1{$^{(##1)}$}                                % Reference in text
[]
  \gdef\refis##1{\indent\hbox to 0pt{\hss##1)~}}        % Ref list numbers)
  \gdef\journal##1, ##2, ##3, 1##4##5##6{               % Journal reference
    {\sl ##1~}{\bf ##2~}(1##4##5##6) ##3}}

\def\pr{\journal Phys. Rev., }

\def\prl{\journal Phys. Rev. Lett., }
\def\prpts{\journal Phys. Rep., }
\def\np{\journal Nucl. Phys., }
\def\pl{\journal Phys. Lett., }

\def\endreferences{\body}
\def\endpage                    %  Eject a page
  {\vfill\eject}
\def\endpaper                   %  Ways to say goodbye
  {\endmode\vfill\supereject}
\def\endit
  {\endpaper\end}
%%      Various user definitions
\def\ref#1{Ref. #1}                     %       for inline references
\def\Ref#1{Ref. #1}                     %       ditto

\def\m@th{\mathsurround=0pt }
\font\twelvesc=cmcsc10 scaled 1200
\def\cite#1{{#1}}
\def\(#1){(\call{#1})}
\def\call#1{{#1}}
\def\taghead#1{}
\def\leaderfill{\leaders\hbox to 1em{\hss.\hss}\hfill}
\def\twiddle{\lower.9ex\rlap{$\kern-.1em\scriptstyle\sim$}}
\def\bigtwiddle{\lower1.ex\rlap{$\sim$}}
\def\gtwid{\mathrel{\raise.3ex\hbox{$>$\kern-.75em\lower1ex\hbox{$\sim$}}}}
\def\ltwid{\mathrel{\raise.3ex\hbox{$<$\kern-.75em\lower1ex\hbox{$\sim$}}}}
\def\square{\kern1pt\vbox{\hrule height 1.2pt\hbox{\vrule width 1.2pt\hskip 3pt
   \vbox{\vskip 6pt}\hskip 3pt\vrule width 0.6pt}\hrule height 0.6pt}\kern1pt}
%%		EQNORDER.TEX			11/05/85	Doug E.
\catcode`@=11
\newcount\tagnumber\tagnumber=0

\immediate\newwrite\eqnfile
\newif\if@qnfile\@qnfilefalse
\def\write@qn#1{}
\def\writenew@qn#1{}
\def\w@rnwrite#1{\write@qn{#1}\message{#1}}
\def\@rrwrite#1{\write@qn{#1}\errmessage{#1}}

\def\taghead#1{\gdef\t@ghead{#1}\global\tagnumber=0}
\def\t@ghead{}

\expandafter\def\csname @qnnum-3\endcsname
  {{\t@ghead\advance\tagnumber by -3\relax\number\tagnumber}}
\expandafter\def\csname @qnnum-2\endcsname
  {{\t@ghead\advance\tagnumber by -2\relax\number\tagnumber}}
\expandafter\def\csname @qnnum-1\endcsname
  {{\t@ghead\advance\tagnumber by -1\relax\number\tagnumber}}
\expandafter\def\csname @qnnum0\endcsname
  {\t@ghead\number\tagnumber}
\expandafter\def\csname @qnnum+1\endcsname
  {{\t@ghead\advance\tagnumber by 1\relax\number\tagnumber}}
\expandafter\def\csname @qnnum+2\endcsname
  {{\t@ghead\advance\tagnumber by 2\relax\number\tagnumber}}
\expandafter\def\csname @qnnum+3\endcsname
  {{\t@ghead\advance\tagnumber by 3\relax\number\tagnumber}}

\def\equationfile{%
  \@qnfiletrue\immediate\openout\eqnfile=\jobname.eqn%
  \def\write@qn##1{\if@qnfile\immediate\write\eqnfile{##1}\fi}
  \def\writenew@qn##1{\if@qnfile\immediate\write\eqnfile
    {\noexpand\tag{##1} = (\t@ghead\number\tagnumber)}\fi}
}

\def\callall#1{\xdef#1##1{#1{\noexpand\call{##1}}}}
\def\call#1{\each@rg\callr@nge{#1}}

\def\each@rg#1#2{{\let\thecsname=#1\expandafter\first@rg#2,\end,}}
\def\first@rg#1,{\thecsname{#1}\apply@rg}
\def\apply@rg#1,{\ifx\end#1\let\next=\relax%
\else,\thecsname{#1}\let\next=\apply@rg\fi\next}

\def\callr@nge#1{\calldor@nge#1-\end-}
\def\callr@ngeat#1\end-{#1}
\def\calldor@nge#1-#2-{\ifx\end#2\@qneatspace#1 %
  \else\calll@@p{#1}{#2}\callr@ngeat\fi}
\def\calll@@p#1#2{\ifnum#1>#2{\@rrwrite{Equation range #1-#2\space is bad.}
\errhelp{If you call a series of equations by the notation M-N, then M and
N must be integers, and N must be greater than or equal to M.}}\else%
 {\count0=#1\count1=#2\advance\count1
by1\relax\expandafter\@qncall\the\count0,%
  \loop\advance\count0 by1\relax%
    \ifnum\count0<\count1,\expandafter\@qncall\the\count0,%
  \repeat}\fi}

\def\@qneatspace#1#2 {\@qncall#1#2,}
\def\@qncall#1,{\ifunc@lled{#1}{\def\next{#1}\ifx\next\empty\else
  \w@rnwrite{Equation number \noexpand\(>>#1<<) has not been defined yet.}
  >>#1<<\fi}\else\csname @qnnum#1\endcsname\fi}

\let\eqnono=\eqno
\def\eqno(#1){\tag#1}
\def\tag#1$${\eqnono(\displayt@g#1 )$$}

\def\aligntag#1\endaligntag
  $${\gdef\tag##1\\{&(##1 )\cr}\eqalignno{#1\\}$$
  \gdef\tag##1$${\eqnono(\displayt@g##1 )$$}}

\def\eqalignno#1{\displ@y \tabskip\centering
  \halign to\displaywidth{\hfil$\displaystyle{##}$\tabskip\z@skip
    &$\displaystyle{{}##}$\hfil\tabskip\centering
    &\llap{$\displayt@gpar##$}\tabskip\z@skip\crcr
    #1\crcr}}

\def\displayt@gpar(#1){(\displayt@g#1 )}

\def\displayt@g#1 {\rm\ifunc@lled{#1}\global\advance\tagnumber by1
        {\def\next{#1}\ifx\next\empty\else\expandafter
        \xdef\csname @qnnum#1\endcsname{\t@ghead\number\tagnumber}\fi}%
  \writenew@qn{#1}\t@ghead\number\tagnumber\else
        {\edef\next{\t@ghead\number\tagnumber}%
        \expandafter\ifx\csname @qnnum#1\endcsname\next\else
        \w@rnwrite{Equation \noexpand\tag{#1} is a duplicate number.}\fi}%
  \csname @qnnum#1\endcsname\fi}

\def\ifunc@lled#1{\expandafter\ifx\csname @qnnum#1\endcsname\relax}

\let\@qnend=\end\gdef\end{\if@qnfile
\immediate\write16{Equation numbers written on []\jobname.EQN.}\fi\@qnend}

\catcode`@=12
%%%%%%%%%%%%%%%%%%%%             REFORDER.TEX              %%%%%%%%%%%%%%%%%%%%
\refstyleprnp
\catcode`@=11
\newcount\r@fcount \r@fcount=0
\def\refreset{\global\r@fcount=0}
\newcount\r@fcurr
\immediate\newwrite\reffile
\newif\ifr@ffile\r@ffilefalse
\def\w@rnwrite#1{\ifr@ffile\immediate\write\reffile{#1}\fi\message{#1}}

\def\writer@f#1>>{}
\def\referencefile{%			  Stuff to write .REF file
  \r@ffiletrue\immediate\openout\reffile=\jobname.ref%
  \def\writer@f##1>>{\ifr@ffile\immediate\write\reffile%
    {\noexpand\refis{##1} = \csname r@fnum##1\endcsname = %
     \expandafter\expandafter\expandafter\strip@t\expandafter%
     \meaning\csname r@ftext\csname r@fnum##1\endcsname\endcsname}\fi}%
  \def\strip@t##1>>{}}

\def\citeall#1{\xdef#1##1{#1{\noexpand\cite{##1}}}}
\def\cite#1{\each@rg\citer@nge{#1}}	% Variable No. of args, separated by ","

\def\each@rg#1#2{{\let\thecsname=#1\expandafter\first@rg#2,\end,}}
\def\first@rg#1,{\thecsname{#1}\apply@rg}	% each@ag is a general purpose
\def\apply@rg#1,{\ifx\end#1\let\next=\relax%	  variable no. of arg. macro.
\else,\thecsname{#1}\let\next=\apply@rg\fi\next}% args separated by commas

\def\citer@nge#1{\citedor@nge#1-\end-}	% Check for M-N range (M and N numbers)
\def\citer@ngeat#1\end-{#1}
\def\citedor@nge#1-#2-{\ifx\end#2\r@featspace#1 % Single argument
  \else\citel@@p{#1}{#2}\citer@ngeat\fi}	% M-N range of arguments
\def\citel@@p#1#2{\ifnum#1>#2{\errmessage{Reference range #1-#2\space is bad.}%
    \errhelp{If you cite a series of references by the notation M-N, then M and
    N must be integers, and N must be greater than or equal to M.}}\else%
 {\count0=#1\count1=#2\advance\count1
by1\relax\expandafter\r@fcite\the\count0,%
  \loop\advance\count0 by1\relax%	  Loop from M to N
    \ifnum\count0<\count1,\expandafter\r@fcite\the\count0,%
  \repeat}\fi}

\def\r@featspace#1#2 {\r@fcite#1#2,}	% Eat spaces at beginning or end of arg
\def\r@fcite#1,{\ifuncit@d{#1}%		  Cite individual reference
    \newr@f{#1}%
    \expandafter\gdef\csname r@ftext\number\r@fcount\endcsname%
                     {\message{Reference #1 to be supplied.}%
                      \writer@f#1>>#1 to be supplied.\par}%
 \fi%
 \csname r@fnum#1\endcsname}
\def\ifuncit@d#1{\expandafter\ifx\csname r@fnum#1\endcsname\relax}%
\def\newr@f#1{\global\advance\r@fcount by1%
    \expandafter\xdef\csname r@fnum#1\endcsname{\number\r@fcount}}

\let\r@fis=\refis			% Save old \refis, redefine
\def\refis#1#2#3\par{\ifuncit@d{#1}%      Use two params #2 #3 to strip blank
   \newr@f{#1}%
   \w@rnwrite{Reference #1=\number\r@fcount\space is not cited up to now.}\fi%
  \expandafter\gdef\csname r@ftext\csname r@fnum#1\endcsname\endcsname%
  {\writer@f#1>>#2#3\par}}

\def\ignoreuncited{%   redefine \refis if ignoring uncited references
   \def\refis##1##2##3\par{\ifuncit@d{##1}%
     \else\expandafter\gdef\csname r@ftext\csname
r@fnum##1\endcsname\endcsname%
     {\writer@f##1>>##2##3\par}\fi}}

\def\r@ferr{\endreferences\errmessage{I was expecting to see
\noexpand\endreferences before now;  I have inserted it here.}}
\let\r@ferences=\references
\def\references{\r@ferences\def\endmode{\r@ferr\par\endgroup}}

\let\endr@ferences=\endreferences
\def\endreferences{\r@fcurr=0%		  Save old \endreferences, redefine
  {\loop\ifnum\r@fcurr<\r@fcount%	  Loop over refnum and produce text
    \advance\r@fcurr by 1\relax\expandafter\r@fis\expandafter{\number\r@fcurr}%
    \csname r@ftext\number\r@fcurr\endcsname%
  \repeat}\gdef\r@ferr{}\global\r@fcount=0\endr@ferences}

\let\r@fend=\endpaper\gdef\endpaper{\ifr@ffile
\immediate\write16{Cross References written on []\jobname.REF.}\fi\r@fend}

\catcode`@=12

\citeall\refto		% These macros will generate citations
\citeall\ref		%
\citeall\Ref		%

\referencefile

\font\titlefont=cmr10 scaled\magstep3
\def\bigtitle                      %  Title on title page
  {\null\vskip 3pt plus 0.2fill
   \beginlinemode \doublespace \raggedcenter \titlefont}

%MACROS FOR KLEIN SYMPOSIUM
%1st updating with effect from: 5 Sept 1991
%2ND UPDATING WITH EFFECT FROM: 28 JUNE 1993
%(for the purpose of making PlainTex file + Latex file identical)

%------------------------------------------------------------------------
\headline={\ifnum\pageno=1\firstheadline\else
\ifodd\pageno\rightheadline \else\leftheadline\fi\fi}
\def\firstheadline{\hfil}
\def\rightheadline{\hfil}
\def\leftheadline{\hfil}
	\footline={\ifnum\pageno=1\firstfootline\else\otherfootline\fi}
\def\firstfootline{\rm\hss\folio\hss}
\def\otherfootline{\hfil}

\font\twelvebf=cmbx10 scaled\magstep 1
\font\twelverm=cmr10 scaled\magstep 1
\font\twelveit=cmti10 scaled\magstep 1

\font\tenbf=cmbx10
\font\tenrm=cmr10
\font\tenit=cmti10

\parindent=1.5pc
\hsize=6.0truein
\vsize=8.5truein
\nopagenumbers

%%%%%%%%%%%%%%%%%%%%%%%%%%%%%%%%%%%%%%%%%%%%%%%%%%%%%%%%%%%%%%%%%%%%%%%%%%%%%%

\centerline{\tenbf PROBING FOR THE ROOTS OF THE STANDARD MODEL
}
\baselineskip=22pt

\vglue 0.8cm
\centerline{\tenrm P.RAMOND}
\baselineskip=13pt
\centerline{\tenit Institute for Fundamental Theory,
 University of Florida}
\baselineskip=12pt
\centerline{\tenit  Gainesville, FL 32611, United States}
\vglue 0.8cm
\centerline{\tenrm ABSTRACT}
\vglue 0.3cm
{\rightskip=3pc
 \leftskip=3pc
 \tenrm\baselineskip=12pt\noindent

The differences between the $N=0$ and $N=1$ standard models are
emphasized in formulating their short distance extension.
We sketch  methods to reproduce  many of the small numbers in
the model in terms of scale ratios, applying see-saw like ideas to the
breaking of chiral symmetries. We sketch how the $N=1$ standard model,
outfitted with an extra family Abelian symmetry to reproduce the mass
hierarchies, naturally fits superstring models, by making use of generic
non-renormalizable operators.
\vglue 0.6cm}
\vfil
\twelverm\baselineskip=14pt
\leftline{\twelvebf 1. Introduction }
\vglue 1pt

In 1938, in one of his many remarkable insights, Oskar Klein
started formulating what later came to be known as a Yang-Mills
theory. Today, we know such theories provide the theoretical
scheme for the interactions of the building blocks of matter,
encapsulated in the Standard Model, in a remarkably compact
description in terms of three gauge groups and eighteen
parameters. Yet, it hardly looks like a
fundamental theory: it has too many unconnected parts, but with
enough similarity among them to lend credence to the belief
that they probably are the chiral shards of a beautiful, more
symmetric underlying structure. We need another Klein to point
us in the right direction. If the Standard Model is indeed  an
effective low energy theory, it must come with  an ultraviolet
cut-off. The {\it raison d'\^ etre} and the value of this
cut-off are the central question of fundamental theory. The
absence of any experimental indication of its existence,
indicates it must be at least of the order of hundreds of GeVs.
At the higher end, Nature provides us with its own cut-off, the
Planck scale, the largest cut-off we can presently imagine to
the standard model. However it is far removed from present
experimental scales.

Local field theories of gravity also break down at the Planck
scale: their generalization to superstring theories may provide
a cure by offering a well-manged deviation from space-time
locality. One, the heterotic string remarkably contains the
basic ingredients needed to reproduce the low energy world.
Then, how do we match the standard model to string theory? One
obvious obstacle is the disparity of scales: the standard model
is known at or below hundreds of GeV's, and string theory
operates in the Planck region, seventeen orders of magnitude
removed. It is quite possible that knowledge of the standard
model alone, may not be sufficient to identify this  match. In
spite of a singular lack of uniqueness, low energy theories
derived from superstrings show many generic features, which may
survive in some form down to experimental energies. A
spectacular feature is the presence of chiral matter. Another
is the possibility of supersymmetry at experimental scales. A
more detailed consequence is the presence of a number of
vector-like particles, some with electroweak quantum numbers,
but with hitherto undetermined $\Delta I_W=0$ masses. Another
is the existence of special non-renormalizable terms, used to
compensate for anomalies in the low energy theory.

In order to  probe for the cut-off of the standard model, we
need to conduct experiments at higher energies. The present
rate of progress is one order of magnitude per human
generation. Unless we find a way to prolong human life, this
method cannot be satisfactory for any one physicist. Theorists,
on the other hand, unimpaired by technical details, can perform
gedanken experiments which are much cheaper, and provide more
immediate, albeit less believable,  results. One tool for this
theoretical journey across the scales is the renormalization
group. With it we can continue the standard model to higher
energies. If its extrapolated parameters show unreasonable
behavior at some scale, it means that we have reached the
cut-off. The question of interest is simply the following: what
is the value of this cut-off? Is it just around the corner, is
it at Planck scale, or...?

We start with a  review of the standard model, and present
arguments for extending its validity to much higher energies.
We then discuss its supersymmetric extension, the $N=1$
standard model, which  is perturbative all the way to the
Planck scale, where we can hope to match it with superstring
models. We  then argue, based on our  knowledge of the Yukawa
sector,  that  certain types of non-renormalizable terms,
generic to superstrings, are needed to understand the pattern
of quark and lepton masses.

\vglue 0.6cm
\leftline{\twelvebf 2. The $N=0$ Standard Model}
\vglue 0.4cm

The $N=0$ standard model is described
by three Yang-Mills groups, each with its own dimensionless gauge
coupling, $\alpha_1$ for the hypercharge $U(1)$, $\alpha_3$ for  QCD,
and $\alpha_2$ for the   weak isospin $SU(2)$. QCD itself predicts
strong CP violation, with strength proportional to a fourth
dimensionless parameter $\overline\theta$.

The electroweak symmetry breaking Higgs sector contains two unknowns,
a dimensionless Higgs self-coupling, and the Higgs mass. The
$``$measured" value of  the Fermi coupling accounts for one parameter,
and the other is the  value of the Higgs mass, one of the two
parameters of the model yet to be determined from experiment. The
Yukawa interactions between the fermions and the Higgs yields the nine
masses of the elementary fermions. This sector also contains three
mixing angles which monitor interfamily decays, and one phase which
describes CP violation.

Two of these parameters have dimensions, the Higgs mass, and the QCD
confinement scale, obtained from $\alpha_3$ by dimensional
transmutation. The QCD scale is a tiny number in Planck units
${\Lambda_{QCD}}\sim 10^{-20} M_{Pl}\ .$ This small number has a
natural  explanation due to  the logarithmic variation  of the QCD
coupling with scale.

The Higgs mass is unknown, although  the electroweak order parameter
is determined by the Fermi constant. In terms of the Planck mass it is
also  very small ${G_F^{-1/2}}\sim 10^{-17} M_{Pl}^{}\ .$ The origin
of this small number is a matter of much speculation. In perturbation
theory the Higgs mass is of the same order of magnitude as the
electroweak order parameter. The most natural idea is to relate this
number to dimensional transmutation associated with new strong
technicolor forces just beyond electroweak scales, so that the natural
cut-off of the standard model is in the TeV range. This beautiful
idea  yields a satisfying natural explanation of this value, but
fails to reproduce the values of the fermion masses.

Another class of extension of the standard model postulates
supersymmetry\refto{reviews} at TeV scales. There, the electroweak
order parameter is related to that  of supersymmetry breaking. While
not at first sight  very economical, the breaking of  supersymmetry
automatically generates electroweak breaking\refto{trigger} in a wide
class of theories. The beautiful ideas of technicolor can then be
applied to supersymmetry breaking, without encountering the problem of
fermion masses of technicolor applied to electroweak breaking.

Whatever the extension, there are many other numbers to explain,
notably in the Yukawa sector of the theory. Quark and charged lepton
masses break electroweak symmetry by $\Delta I_W=1/2$, and
$\vert\Delta Y\vert=1$,  the same quantum numbers as  the
electroweak order parameter, which also gives the W-boson its mass. In
this sense charged fermion masses should be of the same order as the W
mass. This is true only for the top  quark, the others are unnaturally
small
$${m_{u,d}\over M_W}\sim {\cal O}(10^{-4})\ ;\qquad
{m_s\over M_W}\sim {\cal O}(10^{-3})\ ;\qquad {m_c\over M_W}
\sim{\cal O}(10^{-2})\ ;
\qquad {m_b\over M_W}\sim .05\ .$$
Similarly for the charged leptons,
$${m_e\over M_W}\sim {\cal O}(10^{-5})\ ;\qquad
{m_\mu\over M_W}\sim {\cal O}(10^{-3})\ ;\qquad {m_\tau
\over M_W}\sim .02\ ,$$
which  range from the tiny to the small.
Neutrino masses are predicted to be exactly zero in the standard model
only because of the global chiral lepton number symmetries. However there is
mounting experimental evidence that neutrinos have masses. In the
absence of new degrees of freedom they are of the Majorana kind, and
break weak isospin by one unit, as  $\Delta I_W=1$.
Direct experimental limits on neutrino masses indicate that they are at most
extremely small:
${m_{\nu_e}}< 10^{-17} M_W\ .$

The values of the three gauge parameters are known to great accuracy
from measurements at low energy, although because of endemic problems
associated with strong QCD, the color  coupling is the least well known.
Given these parameters, we can extrapolate the standard model to shorter
distances, using the renormalization group perturbatively. The most
interesting effect occurs in the extrapolation of the three gauge
couplings. The hypercharge and weak isospin couplings meet at a scale of
$10^{13}$ GeV, with a value $\alpha^{-1}\approx 43$, but at that scale,
the QCD coupling is much larger, $\alpha_3^{-1}\approx 38$. Thus,
although the quantum numbers indicate  possible unification into a
larger non-Abelian group, the gauge coupling do not follow suit in this
naive extrapolation. Historically, before the couplings were measured to
such accuracy, it was believed that all three did indeed unify in the
ultraviolet. In the ultraviolet, the values of these couplings is less
disparate than at experimental scales. Similarly, nothing spectacular
occurs to the Yukawa couplings. For instance, the botton quark and
$\tau$ lepton Yukawa couplings meet around $10^9$ GeV, and part  in the
deeper ultraviolet. The situation is potentially more extreme in the
Higgs sector because of the renormalization group behavior of the Higgs
self coupling\refto{sher}. We can consider two cases, depending on the
value of the Higgs mass. If it is below $135$ GeV\refto{sher2}, the
self-coupling turns negative somewhere below Planck scale. This results
in a loss of perturbation theory, with  a potential  unbounded from
below. Using the recently announced value of the top quark mass,  a
Higgs mass of $120$ GeV  means that  $``$instability" sets in at 1 TeV,
indicating some new physics  at that scale. When operative, this bound
provides a low (with respect to Planck mass) energy cut-off for the
standard model.

If the Higgs mass is above $200$ GeV, its self-coupling rises
dramatically towards its Landau pole at a relatively low energy scale.
It means loss of  perturbative control of the theory, and sets an upper
bound on the Higgs mass since there is no evidence of any strong
electroweak coupling at experimental scales.  Strong coupling
implies the Higgs is a composite; in the
technicolor scenario it is a condensate of techniquarks.
There is a tiny range of intermediate values for the Higgs mass for
which both the instability and triviality bounds are pushed to scales
beyond the Planck length, and  there is no standard model prediction of
new physics; the cut-off may well be  indeed the Planck scale.

However this is not an entirely satisfactory situation because of the
dependence of the various standard model parameters on the cut-off.
Quantum fluctuations {\it additively} renormalize the Higgs mass with
a term linearly proportional to the cut-off. Thus even if the Higgs
mass is in a region that does not {\it technically} require new
physics below Planck mass, its value is unnaturally small, if Planck
mass is  the cut-off. Only for a low cut-off is its value natural.
Thus we have two possibilities, either expect a low cut-off, or find a
way to alter the cut-off dependence of the electroweak order
parameter. We already know such an example: the cut-off dependence of
any chiral fermion mass is only logarithmic. The reason is chiral
symmetry, which is recovered by setting the fermion mass to zero. It
affords a protection mechanism which  weakens its  cut-off dependence.

\vglue 0.6cm
\leftline{\twelvebf 3. The $N=1$ Standard Model}
\vglue 0.4cm
Supersymmetry avoids the {\it technical}
naturalness problem by
linking any fermion to a boson of the same mass. With exact
supersymmetry, the boson mass finds itself  protected by the chiral symmetry
of the fermion. As long as supersymmetry is broken at energies in the
range of TeV, this is enough protection to produce a low Higgs mass.
This might seem to be small progress, since a new symmetry has been
introduced to relax the strong cut-off dependence, a  symmetry which has to
be broken itself at a small scale, ${V^{}_{SUSY}}\sim 10^{-15} M_{Pl}\ .$

In the $N=1$ standard model, there are only gauge and Yukawa coupling
constants, and their values at experimental scales are such that none
blow up below Planck mass. In particular, the perky Higgs
self-coupling is replaced by the square of gauge and Yukawa couplings.
The great theoretical advantage of the $N=1$ standard model is to
allow the perturbative extrapolation all the way to Planck scale,
opening the way for comparison with fundamental theory!

There are tantalizing hints of simplicity in the extrapolation of the
couplings. Firstly the gauge couplings seem to be much
closer to unification, and at a scale large enough not to be  invalidated
by proton decay
bounds. The hypercharge and weak isospin couplings meet
at a scale of the order of $10^{16}$ GeV, with a value
$\alpha^{-1}\approx 25$, and the QCD coupling is much
closer to, if not right on the same value\refto{unification}. It may
still be a shade higher than the others, with
$(\alpha^{-1}-\alpha_3^{-1})\le 1.5$.

The second remarkable thing is that with simple boundary conditions at
or near Planck mass, inspired by universal soft supersymmetry
breaking, the renormalization group drives one of the Higgs masses to
imaginary values in the infrared. This in turns triggers electroweak
breaking\refto{trigger}, made possible by the large top quark mass!

The Higgs mass is not arbitrarily high in the minimal supersymmetric
extension. At tree-level, it is predicted to be below the Z-mass, but
it suffers large radiative corrections due to the top Yukawa coupling,
raising it above the Z, but not by an arbitrarily large
amount\refto{gordy}.

This general scheme allows us to study the pattern of fermion masses
at these shorter distances; there are more regularities with
supersymmetry. For instance, the bottom quark and $\tau$ masses seem
to unify at or around $10^{16-17}$ GeV\refto{btau}, the same scale
where the gauge couplings converge.

The most striking aspect of the fermion masses is that only one
chiral family has large masses, leading us  to consider theories
where the tree-level Yukawa matrices are simply of the form
$${\bf Y}_{u,d,e}=y^{}_{t,b,\tau}\pmatrix{0&0&0\cr 0&0&0\cr
0&0&1\cr}\ .$$
These matrices imply a global chiral symmetry, $U(2)_L\times U(2)_R$,
in each charged sector. The hierarchy between the bottom and top quark
masses  requires  explanation. In the $N=1$ model, it is linked  to
another parameter which comes from the Higgs sector, the ratio of the
{\it vev} of the two Higgs. Hence it may not pertain to properties of
the Yukawa matrices. Why are the other two families so light? Starting
from the rank two Yukawa matrices, we must find a scheme by which the
zeros get filled, presumably in higher orders of perturbation theory.
In order to see how this might come about, let us examine one
well-known case in which small numbers are naturally generated, the
see-saw mechanism\refto{gmrsy}.

In the standard model, the neutrino Majorana mass matrix is zero at
tree-level. A detailed examination shows that these zeros are
protected from quantum corrections by conservation of  chiral global
lepton number for each species. In the see-saw mechanism,  the usual
neutrinos are mixed with new electroweak singlet fields (neutral
leptons), by $\Delta I_w=1/2$ terms, of electroweak breaking strength,
which  gives them the same lepton numbers. These new particles are
free to acquire $\Delta I_W=0$ Majorana masses, $M$, of any magnitude,
in particular well above the electroweak scale.  Upon diagonalization,
this generates a mass for the familiar neutrinos, depressed from
typical  electroweak values by the ratio of scale ${m\over M}$, where
$m$ is the   electroweak order parameter. A scale ratio between
electroweak and  chiral lepton number  breaking is used to generate a
small number.

We can imagine a similar analysis for the charged Yukawa matrices,
where the zeros are also protected by chiral symmetries. We first
couple the massless fermions of the first two families to  new
fermions with similar quantum numbers, thereby sharing  with them the
chiral symmetries. Unlike the neutral case, these new fermions have
electroweak charges, and can only have Dirac masses, which break the
chiral symmetry. However Dirac masses require  vector-like partners
(this differs from the neutral sector). Thus it may well be that the
small numbers in the mass ratios are to be explained in terms of scale
ratios, underlying the need for extrapolating the standard model many
orders of magnitude. We would also require vector-like particles with
$\Delta I_W=0$ masses, at the higher scales.

The order parameter of supersymmetry breaking is the mass difference
between a particle and its superpartner. For supersymmetry to make sense
in the context of the standard model, this split has to be of the order
of a TeV. How does this come about? It may be useful to draw on an
analogy with chiral symmetry which, while broken by masses, is still an
extraordinarily useful concept.

Chiral symmetry is spontaneously broken both by the strong interactions,
and by electroweak breaking. The latter induces masses for the quarks.
In the limited context of the effective low energy theory of the
Strong Interactions, this breaking may be viewed as an explicit
breaking, in the sense it gives a mass to the associated
Nambu-Goldstone boson.

We can only speculate on the nature of the exact mechanism that causes
supersymmetry breaking. A popular mechanism is akin to technicolor
where a hidden strong color forms a gaugino condensate, which in
supergravity, breaks supersymmetry spontaneously. However, in the
context of the effective low energy theory, this breaking can be
parametrized by explicit breaking terms.

Chiral symmetry starts with massless pseudoscalar mesons (from the
its spontaneous breaking by QCD). They decouple from matter at zero
momentum (low energy theorems). Supersymmetry starts with massless
superparticles, such as gluinos, etc... . They couple to matter in a
manner predetermined by the supersymmetry.

Explicit chiral symmetry breaking is introduced in the chiral
Lagrangean by soft mass terms. Their effect is to give masses to the
pions, and the general covariance properties of these soft breaking
terms is reflected in measureable sum rules among the pseudoscalars,
such as the Gell-Mann-Okubo formula. Similarly, supersymmetry breaking
is introduced in the $N=1$ standard model Lagrangean by soft terms.
These give masses to all the superpartners. However the precise form
of these terms is a question subject to experimental test, for it will
imply sum rules among the superpartners. After a number of
measurements equal to the number of soft parameters, the theory
becomes predictive. Not only the idea of supersymmetry, but also the
precise form of its breaking will be tested at the LHC and the NLC.

In the case of chiral symmetry, we understand the explicit breaking
terms as coming from quark condensates caused by QCD. In
supersymmetry, we may hope to arrive at an equally simple
understanding. Perhaps gaugino condensates may form in a hidden sector
linked to us by the universal gravitational interactions only. In that
case we may expect great simplicity in the breaking terms, yielding
recognizable patterns among the superpartner masses.

The real advantage of low energy supersymmetry is to allow for an
extrapolation of the standard
model to Planck scale. This raises the hope of matching it to a more
fundamental theory. Of these, none is as beautiful as superstring
theory. Thus we limit our discussion to some of its generic features
which might be of use in the phenomenological discussion that will
take place in the next decade.
\vglue 0.6cm
\leftline{\twelvebf 4. Beyond to Superstrings?}
\vglue 0.4cm
Superstring theories provide us, in the words of S. Fubini, with a
glimse of 21st century mathematics.  They are not understood with any
depth, but some of their generic features, applied to low energy
theories can be identified. They yield  effective
low energy gauge theories, valid below a scale $M_U$,  related to the gauge
coupling through the formula \refto{kounnas} $$M_U\approx
2.5\sqrt{\alpha^{}_U}\times 10^{18}\> {\rm GeV}\ .$$ With
$M_X=10^{16}$ GeV, and $\alpha_U^{-1} < \alpha^{-1}_X \approx 25$,
this implies that contact with the superstring can be made provided
that ${M_U/M_X}>50$, so that there is  a slight discrepancy with the
apparent gauge unification scale.

Their second feature  is to produce at lower
energies remnants of $\bf 27$ and $\overline{\bf 27}$ representations
of $E_6$, reducing  in the effective low energy theory to  three
chiral families and  many vector-like particles, with similar quantum
numbers, which may be used in see-saw like mechanisms to generate
small numbers in the Yukawa matrices.  It also means many
intermediate thresholds between the supersymmetry and unification
scales, which is expected since the effective gauge group is usually
larger than the standard model's.

A third   feature is the existence of a local $U(1)$ symmetry,
with anomaly cancelled through the Green-Schwarz mechanism. This
symmetry, though broken close to the Planck scale, may be discernable
in the  extrapolated low energy standard model.
Ib\`a\~nez \refto{Ib}has argued that this symmetry can be used to fix
the weak mixing angle in superstring theories.  Following   Ib\`a\~nez
and Ross\refto{IR}, we argue\refto{US} that this Abelian symmetry
sets the dimensions  of the  Froggatt and Nielsen\refto{FN} Yukawa
operators.

Are any of these features present in the extrapolated low
energy theory? Consider first the unification of the gauge couplings.
It is predicated on two assumptions: that the
weak hypercharge coupling is normalized to its unification into a higher rank
Lie group, such as $SU(5)$, $SO(10)$ or $E_6$\refto{gut}, and on the absence
of  intermediate thresholds with matter carrying strong or
electroweak quantum numbers between $1$ TeV and $10^{16}$ GeV.
The gauge couplings may not exactly unify at $M_X$, and
we may want to alter this simple picture by requiring
 at least one intermediate threshold between the SUSY
scale and the illusory unification scale at $M_X$ to obtain unification at
the string scale $M_U$\refto{MR}.
At one-loop, the couplings $
\alpha_i^{-1}(t)$ for the three gauge groups, ($i=1,2,3$ for
$U(1)_Y$, $SU(2)_L$, and $SU(3)^c$, respectively)
run with scale according to
$$=\alpha_i^{-1}(t_X)+{b_i\over 2\pi}(t-t_X)\ ,$$
where
$t=\ln (\mu/\mu_0)$, $t_X=\ln (M_X/\mu_0)\ ,$
and $\mu_0$ is an arbitrary reference energy.
For the three families and two Higgs doublets of  the
minimal supersymmetric standard model, we have
$b_1=-{33/ 5}\ ;\  b_2=-1\ ;\ b_3=3\ .$
Since the low energies values of  $\alpha_1$ and $\alpha_2$
are known with the greatest accuracy, we use their trajectories to
define $t_X$ as the scale at which they meet:
$$
\alpha^{-1}_X\equiv\alpha_1^{-1} (t_X) = \alpha_2^{-1} (t_X) \ .
$$
The extrapolated data show that
$\alpha^{-1}_X\approx 25$, with $M_X \approx 10^{16}$ GeV.
We do not assume precisely the
same value for $\alpha_3(t_X)$ at that scale; rather we set
$$
\alpha^{-1}_X = \alpha_3^{-1} (t_X) + \Delta \ .
$$
Present
uncertainties  on the QCD coupling suggest that $$|\Delta | \le 1.5\ .
$$
Suppose there is an intermediate threshold above
supersymmetry at
$$
t_I=\ln({M_I/\mu_0})\ ;\qquad t_I<t_X\ ,
$$
caused by new vector-like
particles with electroweak singlet masses at $M_I$.
Their effect is to alter the $b_i$ coefficients:
$$
b_i\to b_i-\delta_i\ ,~~~~i=1,2,3\ .$$
By requiring unification at $M_U$, we find the constraints
$$
{r\over 14} = {t_U - t_X \over t_U - t_I}\ ,\qquad
{q\over 4} = {t_U - t_X - \pi \Delta/2 \over t_U - t_I}
\>\ ,
$$
written in  terms of
$$q\equiv \delta_3-\delta_2
\qquad {\rm and }\qquad {2\over 5}r\equiv\delta_2-\delta_1\ .$$
For vector-like matter generated from superstrings, $q$ and $r$ are integers.
The value of the gauge coupling at unification is now
$$
\alpha^{-1}_U=\alpha^{-1}_X-{1\over 2 \pi}\left [
\delta_2 (t_U - t_I)+ t_U - t_X \right ]\ .$$
These equations have solutions for  non-exotic matter.
For instance when $\Delta =0.82$ with $r=5$, $q=1$, we get
$$M_U=7.5\times 10^{17} {\rm GeV}\ ;\qquad M_I=4.4\times 10^{12} {\rm
GeV}\ ;\qquad \alpha^{-1}_U=11\ .$$
However most  solutions do not allow large $M_X/M_I$.

In realistic superstring models, the  assumption of one intermediate
scale is probably not justified. For several intermediate thresholds,
by applying these equations repeatedly, we obtain similar equations,
with $q$ and $r$ replaced by average quantities which are no longer
integers.  It might seem rather surprising that in the MSSM the gauge
couplings should appear to be nicely headed for unification at $M_X$,
only to be redirected to a new meeting place at $M_U$, but  the
apparent perverseness of this situation allows us put some non-trivial
constraints on the scenario.

Let us now turn to the last topic, the possibility of an
Abelian gauge symmetry, with anomaly cancelled by the Green-Schwarz
mechanism. This gauged Abelian symmetry can play a role in determining
the dimensions of the entries of the Yukawa matrices\refto{IR,US}.
In an effective low energy theory,  anomalies not cancelled by
particles with masses lower than the cut-off, will require
non-renormalizable terms for cancellation. Thus if we can identify
an anomalous  symmetry in the low energy theory, we can hope to learn
something about the the theory beyond the cut-off.

The  most general Abelian charge that can be assigned to
the particles of the Minimal
Supersymmetric Standard Model,  with $\mu$ term, can be written as
$$X=X_0+X_3+{\sqrt 3}X_8\ ,$$
where $X_0$ is the family independent part, $X_3$ is along $\lambda_3$,
and $X_8$ is along $\lambda_8$.
We set
$$X_{i}^{}=(a^{}_{i},b^{}_{i},c_{i}^{},d_{i}^{},e_{i}^{})\ ,$$
where $i=0,3,8$, and  the entries correspond to the components in the
family space of the fields ${\bf Q}$, $\overline{\bf u}$,
$\overline{\bf d}$, $L$, and $\overline e$, respectively. Both Higgs
doublets have the same zero X-charge, without loss of generality,
since an imbalance can be created by mixing in the hypercharge $Y$.

With the tree-level Yukawa coupling {\it only} to the third family, we
obtain the constraints
$${a^{}_0+b^{}_0\over 3}=2(a_8+b_8)\ ,\ \ {a^{}_0+c^{}_0\over 3}=2(a_8+c_8)
\ ,\ \ {d^{}_0+e^{}_0\over 3}=2(d_8+e_8)\ .$$
The other entries in the Yukawa matrices are much smaller, forbidden by
X symmetry to appear at tree level. We assume they appear in the low
energy effective theory as non-renormalizable operators, and that the
excess X-charge at each of their entries is made up by powers of a
single electroweak singlet field. A typical term would be of the form

$${\bf Q}_i\overline{\bf u}_jH_u\left({\theta\over M}\right)^{n_{ij}}\ ,$$
where $\theta$ is some field with unit X-charge, $M$ is some large
scale, and $n_{ij}$ is needed for X conservation.  In order to produce
a small coefficient, the $i$th and $j$th fermions need to go through a
number of intermediate steps to interact. The larger the number steps,
the larger $n_{ij}$, and the smaller the entry in the effective Yukawa
matrix. This approach was advocated long ago by Froggatt and
Nielsen\refto{FN}. This yields  approximate zeros in the matrices,
creating textures\refto{RRR} . For example, in the charge 2/3 sector,
$$n_{12}=3(a_8+b_8)+a_3-b_3\ .$$
Since $\theta$ may have a large expectation value, it may be
accompanied by its vector-like partner $\overline\theta$, with
opposite charge, showing that the exponents $n_{ij}$ need not be
positive, but if  all the $n_{ij}$ are positive, several interesting
phenomenological consequences follow\refto{US}. First the $n_{ij}$
exponents are not all independent,  resulting in order of magnitude
estimates among the Yukawa matrix elements
$$\eqalign{(Y)_{11}&\sim {(Y)_{13}(Y)_{31}\over (Y)_{33}}\ ,\cr
(Y)_{22}&\sim {(Y)_{23}(Y)_{32}\over (Y)_{33}}\ ,\cr   }
$$
valid for each of the three charge sectors. These relations are
consistent with many of the allowed textures. Another
important consequence is that the X-charge of the determinant in each charge
sector is {\it independent} of the texture coefficients that distinguish
between the two lightest families
$${\rm charge}~{2\over 3}~:~6(a_8+b_8)\equiv U\ ,\
{\rm charge}~-{1\over 3}~:~6(a_8+c_8)\equiv D\ ,\
{\rm charge}~-1~:~6(d_8+e_8)\equiv L\ .$$
Let the value of $\theta\over M$ be a small parameter
$\lambda$. In the simplest case, this parameter would be the same for
all three charge sectors. Then we have
$${ m_dm_sm_b\over m_em_\mu m_\tau}\sim {\cal O}(\lambda^{(D-L)})\ .$$
It is more difficult to compare the up and down sectors in this way
since we do not know the value of $\tan\beta$, which sets the
normalization between the two sectors
$${ m_um_cm_t\over m_dm_sm_b}\sim ({y_t\over y_b})^3\tan^3\beta\times{\cal
O}(\lambda^{(U-D)})\ .$$
Since this ratio is much larger
than one, it means either that $\tan\beta$ is itself large, with $U$
close to $D$, or that $\tan\beta$ is not large, but $D>U$.

It can be shown\refto{US} that with one
electroweak singlet field, the X symmetry has to be anomalous to
reproduce the features of the data. The three chiral families
contribute to the mixed gauge anomalies as follows
$$\eqalign{C_3&=2a_0+b_0+c_0\ ,\cr
C_2&=3a_0+d_0\ ,\cr
C_1&={1\over 3}a_0+{8\over 3}b_0+{2\over 3}
c_0+d_0+2e_0\ .\cr}$$
The subscript denotes the gauge group of the Standard Model, {\it i.e.}
$1\sim U(1)$, $2\sim SU(2)$, and $3\sim SU(3)$.
The X-charge also has a mixed gravitational anomaly, which is simply
the trace of the X-charge,
$$C_g=(6a_0+3b_0+3c_0+2d_0+e_0)+C_g^\prime\ ,$$
where $C_g^\prime$ is the contribution from the particles that do not
appear in the minimal $N=1$ model.
The last anomaly coefficient is that of the X-charge itself, $C_X$,
which is the sum of the cubes of the X-charge.

It was suggested by Ib\` a\~ nez\refto{Ib}, that an anomalous $U(1)$
symmetry, with its anomalies cancelled through the Green-Schwarz
mechanism, is capable of relating the ratio of gauge couplings to the
ratios of anomaly coefficients
$${C_i\over k_i}={C_X\over k_X}={C_g\over k_g}\ ,$$
which relates the Weinberg angle to the anomaly coefficients, without
the use of Grand Unification. The $k_i$ are the Kac-Moody levels;  are
integers for the non-Abelian factors only. The mixed $YXX$ anomaly,
however, must vanish by itself.

We demand that the non-Abelian gauge groups have the same
Kac-Moody levels, which means that
$$C_2=C_3\qquad {\rm or}\qquad  d_0=b_0+c_0-a_0\ .$$
Secondly we require that at or near the unification or string scale,
the Weinberg angle have the value
$$\sin^2\theta_W={3\over 8} \ ,$$
which translates into the further constraint
$$5C_2=3C_1\qquad {\rm or}\qquad e_0=2a_0-b_0\ .$$

These equations are sufficient to infer that $L=D$, which implies,
remarkably enough, that the products of the charged lepton masses is
of the same order of magnitude as that of the down-type
quarks\refto{US}.  This provides a remarkable link between the value
of the weak mixing angle and the ratio of down quark to charged lepton
masses.

This formalism has been used\refto{IR} to generate symmetric
textures, of the kind found to be allowed by experiment \refto{RRR}.
Work is in progress to determine how these equations constrain
possible textures.  One result is that it appear to be difficult to
generate acceptable constraints, without invoking Green-Schwarz
cancellation. In that case, this particular way of generating textures
would require the type of mechanism that is generic to superstrings!

The following examples have shown how several problems with the
standard model might require a superstring explanation. While it is
clearly too soon to claim to have made the connection, we think we are
on the way to asymptotic beauty.

I wish to thank Professors Ulf Lindstr\"om and Lars Brink for their
kind invitation and hospitality during this Centenary Celebration.
This work was supported in part by the United States Department of
Energy under Contract No. DEFG05-86-ER-40272.

\references

\refis{FN} C.~Froggatt and H.~B.~Nielsen \np B147, 277, 1979.

\refis{reviews}
For reviews, see H.~P.~Nilles,  \prpts 110, 1, 1984 and
H.~E.~Haber and G.~L.~Kane, \prpts 117, 75, 1985.

\refis{trigger}
L.~E.~Ib\'a\~nez and  G.~G.~Ross,
\journal Phys.~Lett., 110B, 215, 1982;
K.~Inoue, A.~Kakuto, H.~Komatsu, and S.~Takeshita,
\journal Prog. Theor. Phys., 68, 927, 1982;
L.~Alvarez-Gaum\'e, M.~Claudson, and M.~Wise,
\np B207, 16, 1982;
J.~Ellis, J.~S.~Hagelin, D.~V.~Nanopoulos, and
K.~Tamvakis,
\journal Phys.~Lett., 125B, 275, 1983.

\refis{gut}
J.~C.~Pati and A.~Salam,
\pr D10, 275, 1974;
H.~Georgi and S.~Glashow,
\prl 32, 438, 1974;
H.~Georgi, in {\it Particles and Fields-1974}, edited by C.E.Carlson,
AIP Conference Proceedings No.~23 (American Institute of Physics,
New York, 1975) p.575;
H.~Fritzsch and P.~Minkowski,
\journal Ann.~Phys.~NY, 93, 193, 1975;
F.~G\" ursey, P.~Ramond, and P.~Sikivie,
\pl 60B, 177, 1975.

\refis{btau} H.~Arason, D.~J.~Casta\~no, B.~Keszthelyi, S.~Mikaelian,
E.~J.~Piard, P.~Ramond, and B.~D.~Wright,
\prl 67, 2933, 1991;
A.~Giveon, L.~J.~Hall, and U.~Sarid,
\pl 271B, 138, 1991.

\refis{gordy} G.~L.~Kane, C.~Kolda, and J.~D.~Wells,
\prl 70, 2686, 1993.

\refis{sher}M.~Sher, \prpts 179, 273, 1989.

\refis{RRR}P. Ramond, R.G. Roberts and G. G. Ross, \np B406, 19, 1993.

\refis{Ib}L. Ib\'a\~nez, \pl B303, 55, 1993.

\refis{IR}L. Ib\'a\~nez and G. G. Ross, \pl B332, 100, 1994.

\refis{US} P. Bin\'etruy and P. Ramond, in preparation.

\refis{MR}S. Martin and P. Ramond, in preparation.

\refis{unification}
U.~Amaldi, W.~de Boer, and H.~Furstenau,
\pl B260, 447, 1991;
J.~Ellis, S.~Kelley and D.~Nanopoulos,
\pl 260B, 131, 1991;
P.~Langacker and M.~Luo,
\pr D44, 817, 1991.

\refis{sher2} M. Sher \pl B137, 159, 1993;
addendum {\it ibid} {\bf B331}, 448(1994).     .

\refis{gmrsy}M. Gell-Mann, P. Ramond, and R. Slansky in Sanibel
Talk,
CALT-68-709, Feb 1979, and in {\it Supergravity} (North Holland,
Amsterdam 1979). T. Yanagida, in {\it Proceedings of the Workshop
on Unified Theory and Baryon Number of the Universe}, KEK, Japan,
1979.

\refis{kounnas} J.P. Derendinger, S. Ferrara, C. Kounnas, and F. Zwirner, \np
B372, 145, 1992.

\endreferences\endit\end